\documentclass[twocolumn,floatfix,aps,prb,eqsecnum,superscriptaddress]{revtex4}
\usepackage{graphicx}
\usepackage{amsmath}
\usepackage{amssymb}
\usepackage{bm}

\begin{document}

\title{Wiedemann-Franz-type relation between shot noise and thermal conduction of Majorana surface states in a three-dimensional topological superconductor}
\author{N. V. Gnezdilov}
\affiliation{Instituut-Lorentz, Universiteit Leiden, P.O. Box 9506, 2300 RA Leiden, The Netherlands}
\author{M. Diez}
\affiliation{Instituut-Lorentz, Universiteit Leiden, P.O. Box 9506, 2300 RA Leiden, The Netherlands}
\author{M. J. Pacholski}
\affiliation{Institute of Theoretical Physics, Faculty of Physics, University of Warsaw, ul.\ Pasteura 5, 02--093 Warszawa, Poland}
\author{C. W. J. Beenakker}
\affiliation{Instituut-Lorentz, Universiteit Leiden, P.O. Box 9506, 2300 RA Leiden, The Netherlands}
\date{July 2016}
\begin{abstract}
We compare the thermal conductance $G_{\rm thermal}$ (at temperature $T$) and the electrical shot noise power $P_{\rm shot}$ (at bias voltage $V\gg k_{\rm B}T/e$) of Majorana fermions on the two-dimensional surface of a three-dimensional topological superconductor. We present analytical and numerical calculations to demonstrate that, for a local coupling between the superconductor and metal contacts, $G_{\rm thermal}/P_{\rm shot}= {\cal L}T/eV$ (with ${\cal L}$ the Lorenz number). This relation is ensured by the combination of electron-hole and time-reversal symmetries, irrespective of the microscopics of the surface Hamiltonian, and provides for a purely electrical way to detect the charge-neutral Majorana surface states. A surface of aspect ratio $W/L\gg 1$ has the universal shot-noise power $P_{\rm shot}=(W/L)\times (e^2/h)\times (eV/2\pi)$.  
\end{abstract} 
\maketitle

\section{Introduction}
\label{intro}

Topological superconductors are analogous to topological insulators:\cite{Has10,Qi11,Sat16,Bee16} Both combine an excitation gap in the bulk with gapless states at the surface, without localization by disorder as long as time-reversal symmetry is preserved. However, the nature of the surface excitations is entirely different: In a topological insulator these are Dirac fermions, relativistic electrons or holes of charge $\pm e$, while a topological superconductor has \textit{charge-neutral} Majorana fermions on its surface. A transport experiment that aims to detect the Majorana surface states cannot be as routine as electrical conduction --- the direct analogue for Majorana fermions of the \textit{electrical} conductance of Dirac fermions is the \textit{thermal} conductance $G_{\rm thermal}$. The challenge of low-temperature thermal  measurements is one reason why Majorana surface states have not yet been detected in a transport experiment on candidate materials for topological superconductivity.\cite{Hor10,Sas11,Lev13,Mat16,Asa16}

There exists a purely electrical alternative to thermal detection of Majorana fermions.\cite{Akh11} Particle-hole symmetry enforces that a Majorana fermion at the Fermi level is an equal-weight electron-hole superposition, so while the average charge is zero, the charge fluctuations have a quantized variance of
\begin{equation}
{\rm Var}\,Q=\tfrac{1}{2}(+e)^2+\tfrac{1}{2}(-e)^2=e^2.\label{VarQ}
\end{equation}
Quantum fluctuations of the charge can be detected electrically in a shot noise measurement, and for a single fully transmitted Majorana mode these produce a quantized shot noise power $P_{\rm shot}$ of $\tfrac{1}{2}e^2/h$ per $eV$ of applied bias.\cite{Akh11} (The factor $1/2$ reminds us that a Majorana fermion is ``half a Dirac fermion''.)

\begin{figure}[tb]
\centerline{\includegraphics[width=0.9\linewidth]{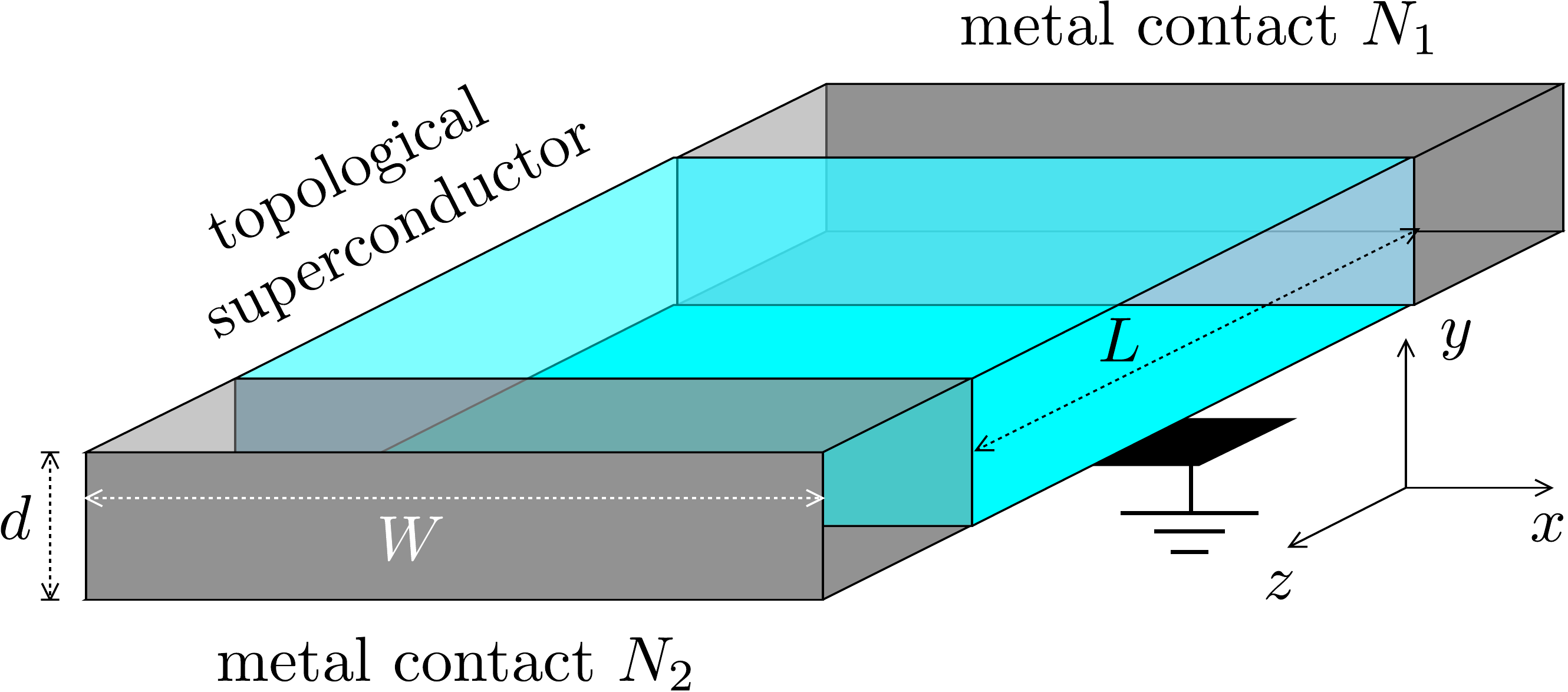}}
\caption{Schematic of a superconductor between a pair of normal-metal contacts. The thermal conductance $G_{\rm thermal}=J/\delta T$ is obtained by applying a temperature difference $T$, $T+\delta T$ between the contacts and measuring the resulting heat current $J$. For the  shot noise measurement one would bias contact $N_1$ at voltage $V$, while keeping  the superconductor and $N_2$ grounded. The electrical current $I_2$ into contact $N_2$ fluctuates with noise power $P_{\rm shot}$. Both $G_{\rm thermal}\propto T$ and $P_{\rm shot}\propto V$ are governed by the Majorana surface states of the topological superconductor.
}
\label{fig_layout}
\end{figure}

In a two-dimensional (2D) topological superconductor, studied in Refs.\ \onlinecite{Die14,Gne15}, there is only a single Majorana edge mode, but a three-dimensional (3D) topological superconductor has a large number $N$ of Majorana surface modes connecting  a pair of metal contacts (see Fig.\ \ref{fig_layout}). In the absence of inter-mode coupling this would give a shot noise power of
\begin{equation}
P_{\rm shot}=\tfrac{1}{2}N{\cal T}P_0,\;\;P_0=eV\frac{e^2}{h},\label{PshotcalT}
\end{equation}
for a mode-averaged transmission probability ${\cal T}$. Because the thermal conductance equals\cite{Sen00}
\begin{equation}
G_{\rm thermal}=\tfrac{1}{2}N{\cal T}G_0,\;\;\;\;G_0={\cal L}T\frac{e^2}{h},\label{GthermalcalT}
\end{equation}
with ${\cal L}=\tfrac{1}{3}(\pi k_{\rm B}/e)^2$ the Lorenz number, uncoupled Majorana modes have a one-to-one relationship between shot noise and thermal conduction.

We would expect this relationship to break down as a result of inter-mode scattering: A pair of coupled Majorana modes is equivalent to a single Dirac fermion mode, which can be in an eigenstate of charge at ${\rm Var}\,Q=0$. The thermal conductance would not be affected, as long as ${\cal T}$ remains the same, but $P_{\rm shot}$ would be reduced. Much to our surprise, we discovered in numerical simulations of a 3D topological superconductor that $P_{\rm shot}/P_0=G_{\rm thermal}/G_0$ with high accuracy. This is remarkable even in the absence of any disorder, since the modes at top and bottom surfaces are coupled when they reach the metal contact. 

We have found that it is the combination of electron-hole and time-reversal symmetry that preserves the relationship between electrical and thermal conduction in a 3D topological superconductor, provided the conversion from Majorana to Dirac fermions at the metal electrode is local in space. The general argument is presented in Secs.\ \ref{conductionformulas} and \ref{PEHTsym}. The implication is that the shot noise power has a universal limit
\begin{equation}
P_{\rm shot}=\frac{1}{2\pi}\frac{W}{L}P_0,\label{Pshotuniversallimit}
\end{equation}
for a surface of aspect ratio $W/L\gg 1$, with corrections from poor coupling to the metal contacts that we calculate in Secs.\ \ref{from3Dto2D} and \ref{sec_Gthermal}. A numerical test of our analytical predictions for a model Hamiltonian of a 3D topological superconductor is given in Sec.\ \ref{sec_3D}. We conclude in Sec.\ \ref{conclude} with a discussion in the context of the Wiedemann-Franz relation between electrical and thermal conduction.\cite{Nom12}

\section{Surface-sensitive thermal and electrical conduction}
\label{conductionformulas}

\subsection{Description of the geometry}
\label{sec_geometry}

We consider the geometry of Fig.\ \ref{fig_layout}, a superconductor $S$ connecting two normal-metal contacts $N_1$ and $N_2$. The superconductor is topologically nontrivial, with a gapped bulk and a gapless surface. We compare two transport properties, one thermal and one electrical, both sensitive to the surface states.

For thermal transport we take the two-terminal thermal conductance
\begin{equation}
G_{\rm thermal}=\lim_{\delta T\rightarrow0}\frac{J}{\delta T},\label{Gthermaldeflinearesponse}
\end{equation}
giving the heat current $J$ flowing from contact $N_1$ at temperature $T+\delta T$ to contact $N_2$ at temperature $T$, in linear response for $\delta T\ll T$. The superconductor is a thermal insulator in the bulk, because of the excitation gap, but a thermal conductor on the surface, so $G_{\rm thermal}$ measures heat conduction along the surface.

For electrical transport both contacts are kept at the same temperature $T$. Contact $N_1$ is biased at voltage $V$ relative to ground, while contact $N_2$ as well as the superconductor are grounded. Most of the charge current $I_1$ injected into the superconductor at $N_1$ is short-circuited to ground via the bulk, which is an ideal electrical conductor. At the remote contact $N_2$ a fluctuating current $I_2(t)$ remains due to surface conduction from $N_1$ to $N_2$. Even if the time average $\langle I_2\rangle$ vanishes, there will be time-dependent fluctuations $\delta I_2(t)$ with low-frequency noise power
\begin{equation}
P_{\rm shot}=\int_{-\infty}^{\infty} dt\,\langle\delta I_2(0)\delta I_2(t)\rangle.\label{Pshotdeflinearresponse}
\end{equation}
At low temperatures $k_{\rm B}T\ll eV$ this is predominantly shot noise $\propto V$.

\subsection{Scattering formulas}
\label{sec_scatteringform}

In a scattering formulation the thermal and electrical transport properties can be expressed in terms of the matrix $t(E)$ of transmission amplitudes from $N_1$ to $N_2$, at energy $E$ relative to the Fermi level. The transmission matrix has a block structure in the electron-hole degree of freedom,
\begin{equation}
t=\begin{pmatrix}
t_{ee}&t_{eh}\\
t_{he}&t_{hh}
\end{pmatrix}.\label{tblock}
\end{equation}
The submatrix $t_{ee}$ describes transmission of an electron as an electron, while $t_{he}$ describes transmission of an electron as a hole.

At sufficiently small temperature and voltage the transmission matrix may be evaluated at the Fermi level ($E=0$) and we have the Landauer-type formulas \cite{Siv86,Ana96,Bee15}
\begin{align}
&G_{\rm thermal}=\tfrac{1}{2}G_0\,{\rm Tr}\,t^\dagger t,\label{Gthermaldef}\\
&P_{\rm shot}=P_0\,{\rm Tr}\,(\tau_+ -\tau_-^2),\label{Pshotdef}\\
&\tau_\pm=t^\dagger_{ee}t^{\vphantom{\dagger}}_{ee}\pm t^\dagger_{he}t^{\vphantom{\dagger}}_{he}.\label{taupmdef}
\end{align}
Eq.\ \eqref{Pshotdef} may equivalently be written in terms of the full transmission matrix, 
\begin{equation}
P_{\rm shot}/P_0=\tfrac{1}{2}\,{\rm Tr}\,(1+\tau_z)t^\dagger t-\tfrac{1}{4}\,{\rm Tr}\,\left[(1+\tau_z) t^\dagger \tau_z t\right]^2,\label{Pshotdef2}
\end{equation}
with the help of the Pauli matrix $\tau_z={{1\;\;\; 0}\choose{\,0\; -1}}$ acting on the electron-hole degree of freedom. 

\subsection{Electron-hole symmetry enforced upper bound on the shot noise power}
\label{sec_symmetries}

Electron-hole symmetry at the Fermi level equates
\begin{equation}
t=\tau_x t^\ast\tau_x,\label{tehsymmetry}
\end{equation}
with $\tau_x={{0\;\; 1}\choose{1\;\;0}}$ the Pauli matrix that exchanges electrons and holes. It follows that 
\begin{align}
{\rm Tr}\,\tau_z t^\dagger t&=\tfrac{1}{2}\,{\rm Tr}\,(t\tau_z t^\dagger+t^\ast\tau_z t^{\rm T})\nonumber\\
&=\tfrac{1}{2}\,{\rm Tr}\,(t\tau_z t^\dagger+\tau_x t\tau_x\tau_z\tau_x t^\dagger\tau_x)=0.\label{trtauztdaggert}
\end{align}

Eqs.\ \eqref{Gthermaldef} and \eqref{Pshotdef2} can therefore be combined into
\begin{equation}
\begin{split}
&P_{\rm shot}/P_0=G_{\rm thermal}/G_0-\delta p,\\
&\delta p=\tfrac{1}{4}\,{\rm Tr}\,\left[(1+\tau_z) t^\dagger \tau_z t\right]^2.
\end{split}
\label{PG0ehsymm}
\end{equation}
Since $(1+\tau_z)^2=2(1+\tau_z)$ the term $\delta p$ can be written as the trace of a positive definite matrix,
\begin{equation}
\delta p={\rm Tr}\,X^2\geq 0,\;\;X=\tfrac{1}{4}(1+\tau_z) t^\dagger \tau_z t(1+\tau_z)=X^\dagger,\label{deltapdef}
\end{equation}
so the dimensionless shot noise power $P_{\rm shot}/P_0$ is bounded from above by the dimensionless thermal conductance $G_{\rm thermal}/G_0$.

In Ref.\ \onlinecite{Akh11} it was demonstrated that this inequality becomes a strict {\em equality} for a rank-one transmission matrix $t$, as in 1D transmission via the unpaired Majorana edge mode of a 2D topological superconductor.\cite{Die14,Gne15} Only particle-hole symmetry is needed in that case. In the next section we will show that the combination of particle-hole symmetry and time-reversal symmetry achieves approximate equality on the 2D surface of a 3D topological superconductor, irrespective of the rank of $t$.

\section{Combined effects of electron-hole and time-reversal symmetries on the shot noise power}
\label{PEHTsym}

\subsection{Surface Hamiltonian with tunnel coupling to metal contacts}
\label{sec_surfaceH}

The surface Hamiltonian of Majorana fermions in the $x$--$z$ plane has the form\cite{Vol03}
\begin{equation}
H=v p_x\sigma_x + v p_z\sigma_z,\label{Hdef}
\end{equation}
with velocity $v$, momentum operators $p_\alpha=-i\partial/\partial x_\alpha$, and Pauli matrices $\sigma_\alpha$ acting on the spin degree of freedom. (The $2\times 2$ unit matrix is $\sigma_0$ and we have set $\hbar$ to unity.) We assume that there is no valley degeneracy of the surface states, so the surface spectrum consists of a single nondegenerate cone with dispersion relation $E^2=v^2 p^2$.

A disorder potential $V(x,z)\sigma_\alpha$ is forbidden by the combination of electron-hole symmetry and time-reversal symmetry, 
\begin{equation}
H=-H^\ast,\;\;H=\sigma_y H^\ast\sigma_y.\label{symmetryH} 
\end{equation}
This insensitivity to impurity scattering is a unique property of a topological superconductor in symmetry class DIII.\cite{Fos14,Que15,note0} A spatial modulation of the Fermi velocity $v(x,z)$ is allowed by symmetry,\cite{Nak14} and this is the only source of scattering on the surface.

The normal metal has propagating modes labeled by a spin degree of freedom $\sigma$, electron-hole degree of freedom $\tau$, and orbital degree of freedom $\nu$. The superconducting surface has only evanescent modes at the Fermi level, because of the vanishing density of states. A Majorana fermion with spin $\sigma'$ at point $\bm{r}$ is coupled to the metal by the coupling matrix element $\langle\sigma',\bm{r}|\Xi|\sigma,\tau,\nu\rangle$. The scattering matrix is given by\cite{Mah69}
\begin{equation}
S(E)=1+i\,\Xi^\dagger(H-\tfrac{1}{2}i\,\Xi\Xi^\dagger -E)^{-1}\Xi,\label{SXidef}
\end{equation}
near the Fermi level where the energy dependence of the coupling matrix $\Xi$ can be neglected.

The scattering matrix is unitary, $SS^\dagger=S^\dagger S=1$, with electron-hole and time-reversal symmetries\cite{note2}
\begin{equation}
S(E)=\tau_x S^\ast(-E)\tau_x,\;\;S(E)=\sigma_y S^{\rm T}(E)\sigma_y.\label{SehTsym}
\end{equation}
The corresponding symmetry relations for the coupling matrix are
\begin{equation}
\Xi=\Xi^\ast\tau_x,\;\;\Xi=\sigma_y\Xi^\ast\sigma_y.\label{XiehTsym}
\end{equation}

\subsection{Condition on the coupling matrix for equality of shot noise and thermal conductance}
\label{conditiononXi}

We now restrict ourselves to the Fermi level, $E=0$, and determine a condition on the coupling matrix $\Xi$ that ensures the equality $P_{\rm shot}/P_0=G_{\rm thermal}/G_0$ of the dimensionless shot noise power and thermal conductance. According to Eqs.\ \eqref{PG0ehsymm} and \eqref{deltapdef}, a necessary and sufficient condition is that the electron-electron block $X$ of the transmission matrix product $t^\dagger\tau_z t$ vanishes identically. Here we establish a \textit{sufficient} condition involving only the coupling matrix:
\begin{equation}
\begin{split}
&\Xi\tau_z \Xi^\dagger\equiv 0\Rightarrow t^\dagger\tau_z t\equiv 0\Rightarrow X\equiv 0\\
&\Leftrightarrow P_{\rm shot}/P_0=G_{\rm thermal}/G_0.
\end{split}
\label{seriesofequivalences}
\end{equation}
No requirements are made on the rank of $t$, which may involve strong mode-mixing by the surface Hamiltonian.

The combination of the electron-hole and time-reversal symmetries \eqref{XiehTsym} implies that
\begin{subequations}
\label{calKsymmetries}
\begin{align}
&\Xi\tau_x=\sigma_y\Xi\sigma_y\Rightarrow\sigma_y\Xi\tau_z \Xi^\dagger=-\Xi\tau_z \Xi^\dagger\sigma_y,\label{calKsymmetriesa}\\
&\Xi\tau_x=\Xi^\ast\Rightarrow \Xi\tau_z \Xi^\dagger=-(\Xi\tau_z \Xi^\dagger)^{\rm T}
,\label{calKsymmetriesb}
\end{align}
\end{subequations}
so the matrix $\Xi\tau_z \Xi^\dagger$ is antisymmetric and it anticommutes with $\sigma_y$. Because the only $2\times 2$ matrix with both these properties is identically zero, we conclude that $\Xi\tau_z\Xi^\dagger$ vanishes if it is block-diagonal in $2\times 2$ matrices:
\begin{equation}
\Xi\tau_z\Xi^\dagger=0\;\;{\rm if}\;\;\langle\sigma',\bm{r}'|\Xi\tau_z\Xi^\dagger|\sigma,\bm{r}\rangle=\Gamma_{\sigma\sigma'}(\bm{r})\delta(\bm{r}-\bm{r}').\label{blockdiagonal}
\end{equation}
The $2\times 2$ matrix $\Gamma(\bm{r})$ acts on the spin degree of freedom of a Majorana fermion at position $\bm{r}$ on the superconducting surface. We will refer to the condition \eqref{blockdiagonal} as a \textit{locality condition} on the coupling matrix product $\Xi\tau_z\Xi^\dagger$.

In the next section we will explicitly solve a simple model where the locality condition holds, but we argue that it is a natural assumption for a weakly disordered NS interface between the normal metal (N) and the topological superconductor (S). If the disorder mean free path is large compared to the superconducting coherence length, a Majorana fermion transmitted through the NS interface is locally converted into an electron-hole superposition via a $2\times 4$ matrix $K$ and then scattered nonlocally in the normal metal via a unitary matrix $U$ without mixing electrons and holes --- so $U$ commutes with $\tau_z$ and $U\tau_z U^\dagger=\tau_z$. Substitution of $\Xi=K U$ gives the desired locality to $\Xi\tau_z\Xi^\dagger=K\tau_z K^\dagger$.

\section{Formulation and solution of the surface scattering problem}
\label{from3Dto2D}

\subsection{Reduction to an effectively 2D geometry}
\label{2Dreduction}

The transmission matrix $t$ refers to a 3D scattering problem, and this is how we will calculate it numerically later on. For an analytical treatment a reduction to an effectively 2D geometry is desirable. Referring to Fig.\ \ref{fig_layout}, in a typical thin-film geometry one has $d\ll W$, so the contributions from the top and bottom surfaces in the $x$--$z$ plane dominate over the contributions from the lateral surfaces in the $y$--$z$ plane.

\begin{figure}[tb]
\centerline{\includegraphics[width=0.5\linewidth]{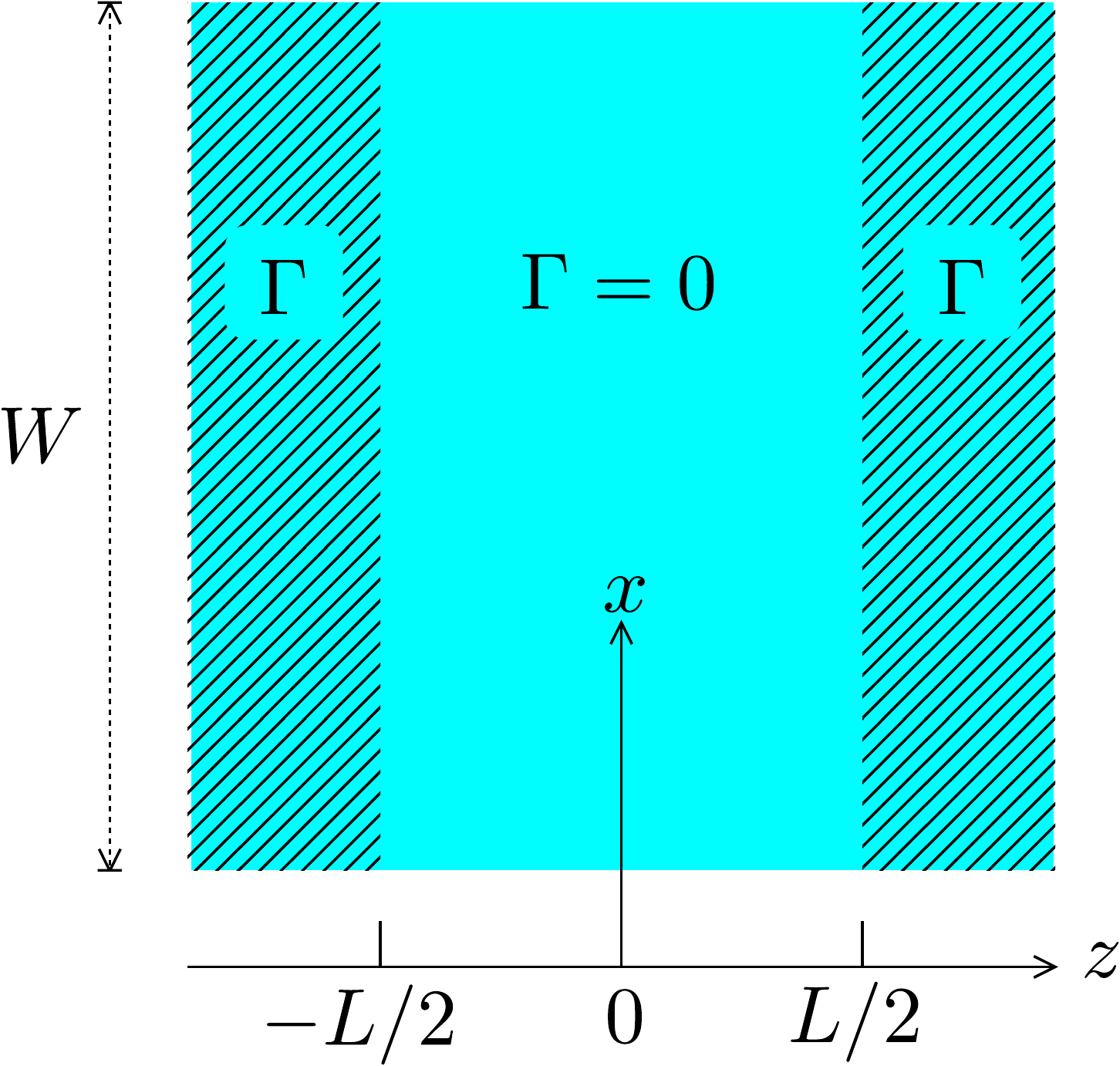}}
\caption{Superconducting surface layer in the $x$--$z$ plane with tunnel coupling to a normal-metal in the region $|z|>L/2$ (shaded, tunnel rate $\Gamma$). This 2D scattering geometry effectively describes the transmission from $N_1$ to $N_2$ via the top or bottom surface of the 3D topological superconductor in Fig.\ \ref{fig_layout}. The extension $d$ of the contact region is assumed to be sufficiently large that top and bottom surfaces can be treated independently. (Coupling of top and bottom surfaces is described in Fig.\ \ref{fig_layout2Dcoupled}.)
}
\label{fig_layout2D}
\end{figure}

The normal-metal contact region in the $x$--$y$ plane is connected to the superconducting surface in the $x$--$z$ plane at $z=\pm L/2$. We ignore the curvature of the surface at this connection and replace the contact region by the region $|z|>L/2$ in the $x$--$z$ plane. Tunneling into the metal electrode in the contact region is described by the effective Hamiltonian
\begin{equation}
{\cal H}=H-\tfrac{1}{2}i\Gamma(z)\sigma_0,\;\;\Gamma(z)=\Gamma\theta(|z|-L/2),\label{tunnelH}
\end{equation}
with $\theta(s)$ the unit step function. If the tunnel barrier is sufficiently transparent (tunnel rate $\Gamma\gg v/d$), a particle approaching $z=\pm L/2$ via the top surface ($y=d/2$) will enter the metal contact before reaching the bottom surface ($y=-d/2$), so that we can treat top and bottom surfaces separately. This produces the effectively 2D geometry of Fig.\ \ref{fig_layout2D}, which we will analyze in the next subsection. The regime $\Gamma\lesssim v/d$, when top and bottom surfaces cannot be treated separately, is considered in Sec.\ \ref{sec_surfacecoupling}.

\subsection{Single-surface transmission matrix}
\label{sec_tmatrix}

The normal metal has a nonzero density of states at the Fermi level, with a set of $M$ transverse momenta $q$ (in the $x$-direction). At each $q$ there are four propagating modes, including the spin and electron-hole degree of freedom. We collect the total number of $4M$ mode indices in the label $\alpha$, with Pauli matrices $\sigma_i$, $\tau_i$ acting, respectively, on the spin and electron-hole degree of freedom. The scattering matrix element $S_{\alpha\alpha'}(z,z';E)$ at energy $E$ relates an outgoing mode $\alpha$ at $z$ to an incoming mode $\alpha'$ at $z'$. 

The full $4M\times 4M$ scattering matrix $S(z,z';E)$ describes both transmission (when $z>L/2$ and $z'<-L/2$ or the other way around) and reflection (when $z,z'>L/2$ or $z,z'<-L/2$). In accord with Eq.\ \eqref{SXidef} it is given by
\begin{equation}
S(z,z';E)=\openone\delta(z-z')+i\Gamma W^\dagger(z) {\cal G}(z,z';E)W(z'),\label{SWGdef}
\end{equation}
in terms of a $2M\times 2M$ matrix Green's function ${\cal G}(z,z';E)$,
\begin{equation}
({\cal H}-E){\cal G}(z,z';E)=\openone\delta(z-z'),\label{GEdef}
\end{equation}
and a $2M\times 4M$ coupling matrix $W(z)$. The rank of the matrix ${\cal G}$ is only half the rank of $S$, because the Majorana fermions on the superconducting surface lack the electron-hole degree of freedom of the Dirac fermions in the normal metal.\cite{note1}

Particle conservation (unitarity) requires that
\begin{equation}
\int_{|z''|>L/2}dz''\, S(z,z'';E)S^{\dagger}(z',z'';E)=\openone\delta(z-z'),\label{unitarity}
\end{equation}
for $|z|,|z'|>L/2$, which is satisfied if 
\begin{equation}
W(z)W^\dagger(z)=\openone.\label{WWdagger}
\end{equation}

Although in the general treatment of the previous section we allowed for mode-mixing on the superconducting surface, for a tractable analytical calculation we now simplify to the Hamiltonian \eqref{Hdef} with a uniform velocity $v$. Because other sources of scattering are excluded by the combination of electron-hole and time-reversal symmetry, the transverse momentum $q$ is not coupled by the effective Hamiltonian \eqref{tunnelH} and ${\cal G}$ decomposes into $2\times 2$ $q$-dependent blocks ${\cal G}(z,z';q,E)$.

This matrix Green's function is calculated in App.\ \ref{app_Gsingle}. To obtain the transmission matrix from $N_1$ to $N_2$ we must take $z>L/2$ and $z'<-L/2$. At the Fermi level the result is
\begin{subequations}
\label{Gkappagamma}
\begin{align}
{\cal G}(z,z';q,0)={}&\frac{1}{2iv}\frac{\exp[-\xi(z-z'-L)]}{\xi\cosh Lq+q\sinh Lq }\nonumber\\
&\times\begin{pmatrix}
-\xi-\kappa& iq\\
iq&\xi-\kappa
\end{pmatrix},\label{Gzeroresult}\\
\xi=\sqrt{q^2+\kappa^2},&\;\;\kappa=\tfrac{1}{2}\Gamma/v,\;\;z>L/2,\;\;z'<-L/2.\label{xikappadef}
\end{align}
\end{subequations}

The $4M\times 4M$ transmission matrix $t(z,z')$ at the Fermi level follows from Eq.\ \eqref{SWGdef},
\begin{equation}
\begin{split}
&t(z,z')=\sum_q t(z,z';q),\\
&t(z,z',q)=i\Gamma w^\dagger(z,q){\cal G}(z,z';q,0)w(z',q),
\end{split}
\label{tWG}
\end{equation}
with a $q$-dependent $2\times 4M$ coupling matrix $w(z,q)$. The unitarity constraint reads
\begin{equation}
w(z,q)w^\dagger(z,q')=\delta_{qq'}\sigma_0.\label{WzqWzqdagger}
\end{equation}

\subsection{Transmission matrix for coupled top and bottom surfaces}
\label{sec_surfacecoupling}

\begin{figure}[tb]
\centerline{\includegraphics[width=0.7\linewidth]{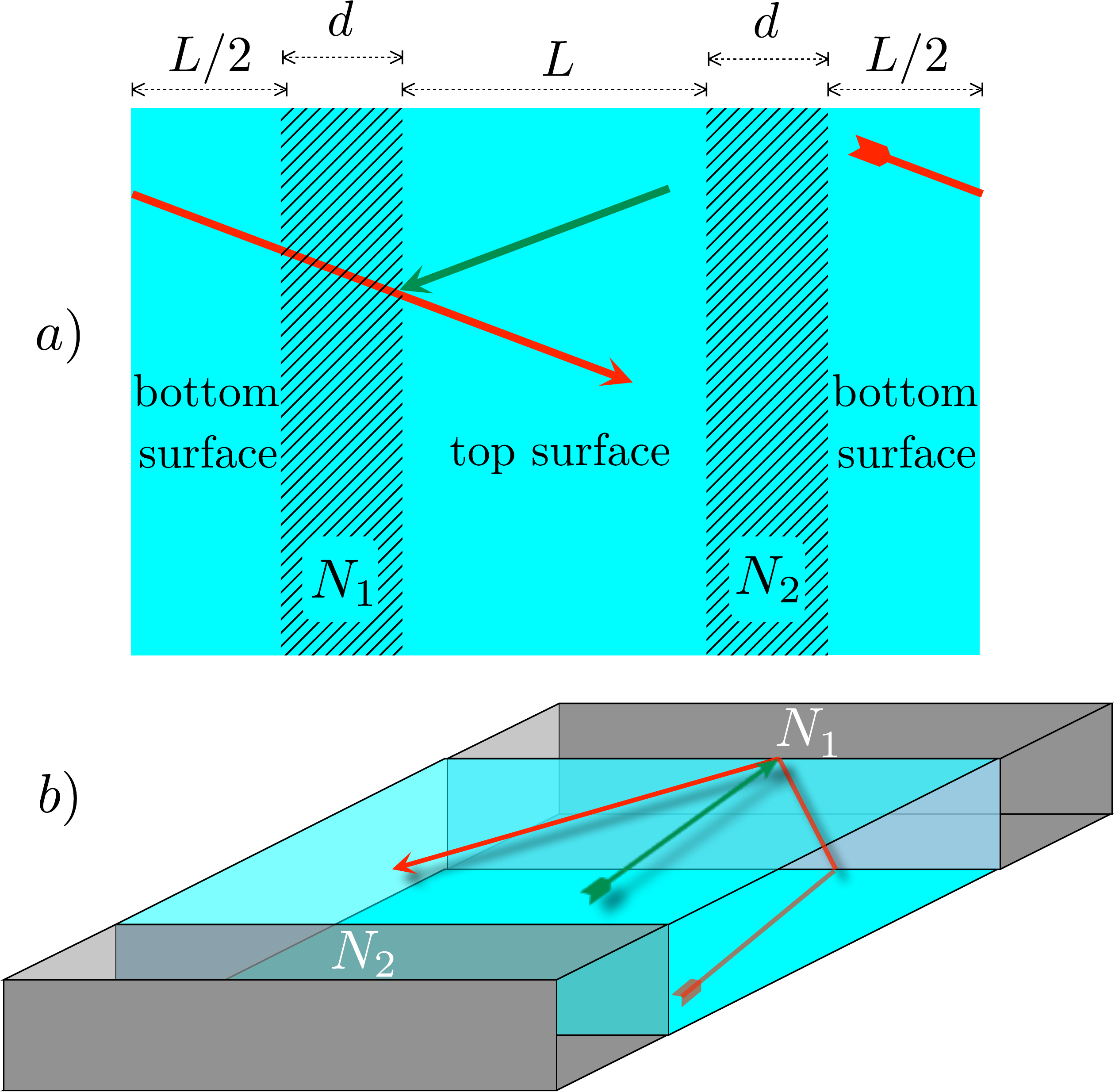}}
\caption{Panel $a$: Same as Fig.\ \ref{fig_layout2D}, but now with antiperiodic boundary conditions at $z=\pm(L+d)$ to include the effect of a coupling of top and bottom surface via the contact region of finite length $d$. Panel $b$ shows the correspondence between trajectories in the 2D and 3D representation. Coupling of the red and green trajectories, at the same transverse momentum $q$, elevates the rank of the $q$-dependent transmission matrix from one to two.
}
\label{fig_layout2Dcoupled}
\end{figure}

The approach outlined above for the case of uncoupled top and bottom surfaces can be readily generalized to allow for a coupling of the two surfaces via the contact region. In the effective 2D representation the region of nonzero $\Gamma$ now extends over a finite interval, 
\begin{equation}
\Gamma(z)=\Gamma\left[\theta(|z|-L/2)-\theta(|z|-L/2-d)\right].\label{Gammazcoupled}
\end{equation}
A particle on the top surface crossing the contact region without being absorbed continues on the bottom surface. The corresponding 2D geometry is shown in Fig.\ \ref{fig_layout2Dcoupled}. It has a finite extent $2L+2d$ in the $z$-direction, with antiperiodic boundary conditions at $z=\pm (L+d)$ to account for a $\pi$ Berry phase.

The calculation of the Green's function is given in App.\ \ref{app_Gdouble}. Instead of Eq.\ \eqref{Gkappagamma} we now have
\begin{subequations}
\label{Gkappagammacoupled}
\begin{align}
&{\cal G}(z,z';q,0)=\frac{1}{2iv}\frac{\cosh\xi s}{\xi\cosh Lq\cosh\xi d+q\sinh Lq\sinh \xi d }\nonumber\\
&\qquad\times\begin{pmatrix}
-\xi-\kappa\tanh \xi s& iq\tanh\xi s\\
iq\tanh\xi s&\xi-\kappa\tanh\xi s
\end{pmatrix},\label{Gzeroresultcoupled}\\
&s=L+d-z+z'\in(-d,d),\nonumber\\
&-L/2-d<z'<-L/2,\;\;L/2<z<L/2+d.\label{szzprimedef}
\end{align}
\end{subequations}
The single-layer result \eqref{Gkappagamma} is recovered in the limit $d\rightarrow\infty$.

The key distinction between the single-surface Green's function \eqref{Gkappagamma} and the coupled-surface result \eqref{Gkappagammacoupled} is that --- while both are $2\times 2$ matrices --- the latter is of rank two but the former is only of rank one (one of the two eigenvalues of the matrix \eqref{Gkappagamma} vanishes).

\section{Results for thermal conductance and corresponding shot noise power}
\label{sec_Gthermal}

We use the 2D surface theory of the previous section to calculate the thermal conductance $G_{\rm thermal}$, including the effects of poor coupling to the metal contacts, strong coupling of top and bottom surfaces, and effects of a finite aspect ratio. Subject to the locality condition \eqref{blockdiagonal} these results apply as well to the shot noise power $P_{\rm shot}=G_{\rm thermal}\times P_0/G_0$.

\subsection{Single surface}
\label{Gthermalsingle}

The thermal conductance \eqref{Gthermaldef} follows from the transmission matrix \eqref{tWG} upon integration,
\begin{equation}
G_{\rm thermal}=\tfrac{1}{2}G_0\int_{L/2}^\infty dz\int_{-\infty}^{-L/2}dz'\,{\rm Tr}\, t^\dagger(z,z')t(z,z').\label{Gthermalt}
\end{equation}
Because of the unitarity condition \eqref{WzqWzqdagger} the coupling matrix drops out and only the $2\times 2$ matrix Green's function enters. Substitution of the result \eqref{Gkappagamma} for a single surface gives
\begin{align}
G_{\rm thermal}/G_0={}&\tfrac{1}{2}\Gamma^2\sum_q\int_{L/2}^\infty dz\int_{-\infty}^{-L/2}dz'\nonumber\\
&\times{\rm Tr}\,{\cal G}^\dagger(z,z';q,0){\cal G}(z,z';q,0)\nonumber\\
={}&\tfrac{1}{2}\kappa^2\sum_q\,(\xi\cosh Lq+q\sinh Lq)^{-2}.\label{Gthermalq}
\end{align}
For $W\gg L$ the sum over transverse momenta may be replaced by an integration, $\sum_q\rightarrow (W/2\pi)\int_{-\infty}^\infty dq$, resulting in
\begin{equation}
G_{\rm thermal}=G_0\frac{W}{2\pi}\int_{0}^\infty dq\,\frac{\kappa^2}{(\xi\cosh qL+q\sinh qL)^2}.\label{Gthermalresult}
\end{equation}

\begin{figure}[tb]
\centerline{\includegraphics[width=0.8\linewidth]{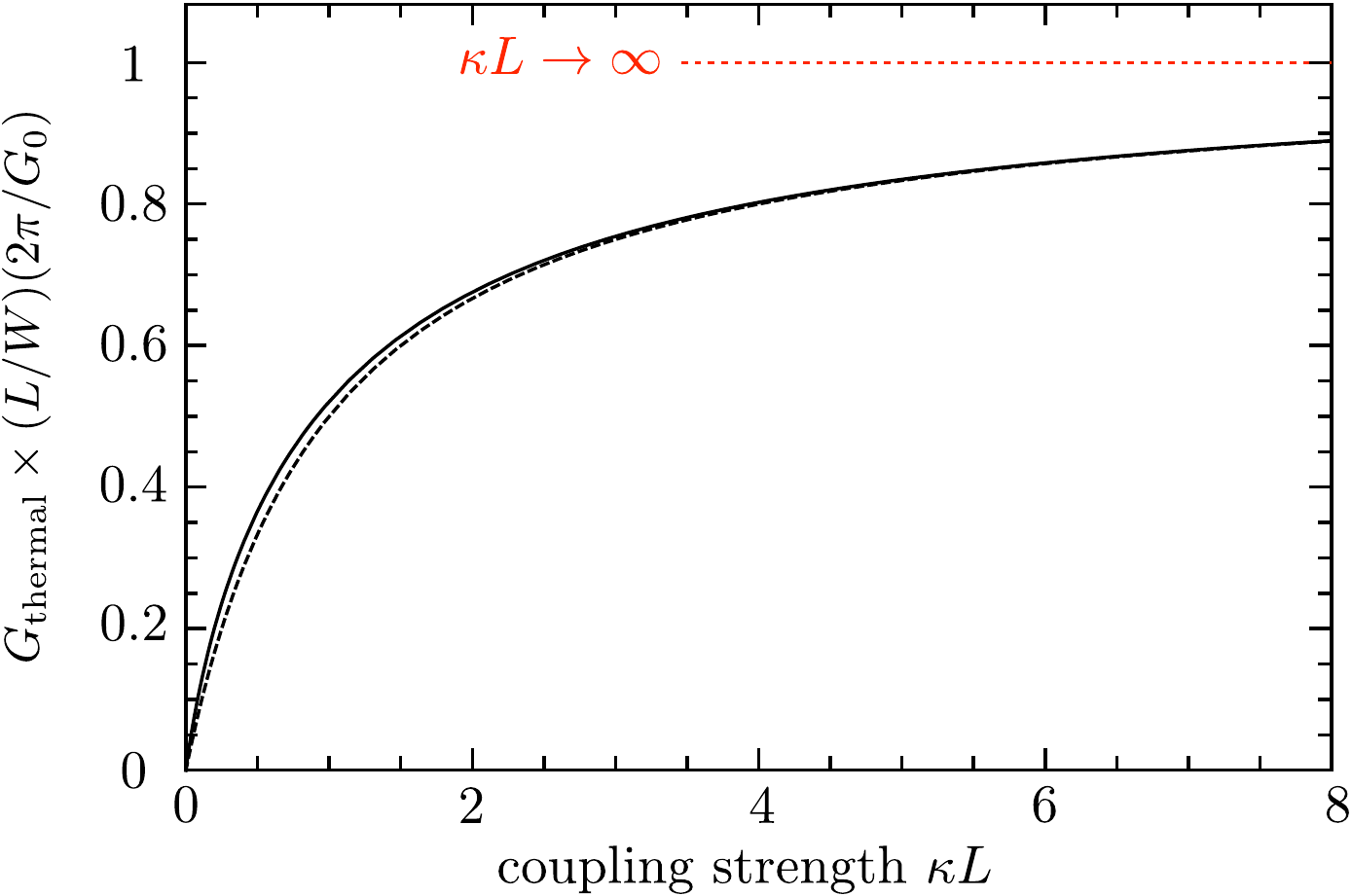}}
\caption{Thermal conductance of the surface of a 3D topological superconductor (aspect ratio $W/L\gg 1$) as a function of the coupling strength to the normal-metal contacts. The solid curve is calculated from Eq.\ \eqref{Gthermalresult}, the dashed curve is the effective-length approximation \eqref{sigmathermalappr}. In the strong-coupling limit $G_{\rm thermal}\rightarrow (W/L)(G_0/2\pi)$.
}
\label{fig_Gthermal}
\end{figure}

The coupling strength of the superconductor to the metal contacts is quantified by the product $\kappa L=\Gamma L/2v$. In the strong-coupling limit $\kappa L\rightarrow\infty$ the thermal conductivity approaches the universal value
\begin{equation}
\frac{L}{W}G_{\rm thermal}\rightarrow \frac{1}{2\pi}G_0,\;\;{\rm for}\;\;\kappa L\rightarrow\infty,\label{sigmathermaluniversal}
\end{equation}
This is the Majorana fermion analogue\cite{Nak14,Xie15} of the Dirac fermion conductivity in graphene.\cite{Kat06,Two06} The effect of a finite coupling strength can be understood in terms of an effective length $L_{\rm eff}=L+1/\kappa$, so that 
\begin{equation}
G_{\rm thermal}\approx \frac{1}{2\pi}G_0\times \frac{W}{L_{\rm eff}}=\frac{G_0 W}{2\pi L}\frac{\kappa L}{1+\kappa L},\label{sigmathermalappr}
\end{equation}
see Fig.\ \ref{fig_Gthermal}.

\subsection{Coupled surfaces}
\label{Gthermalcoupled}

\begin{figure}[tb]
\centerline{\includegraphics[width=0.8\linewidth]{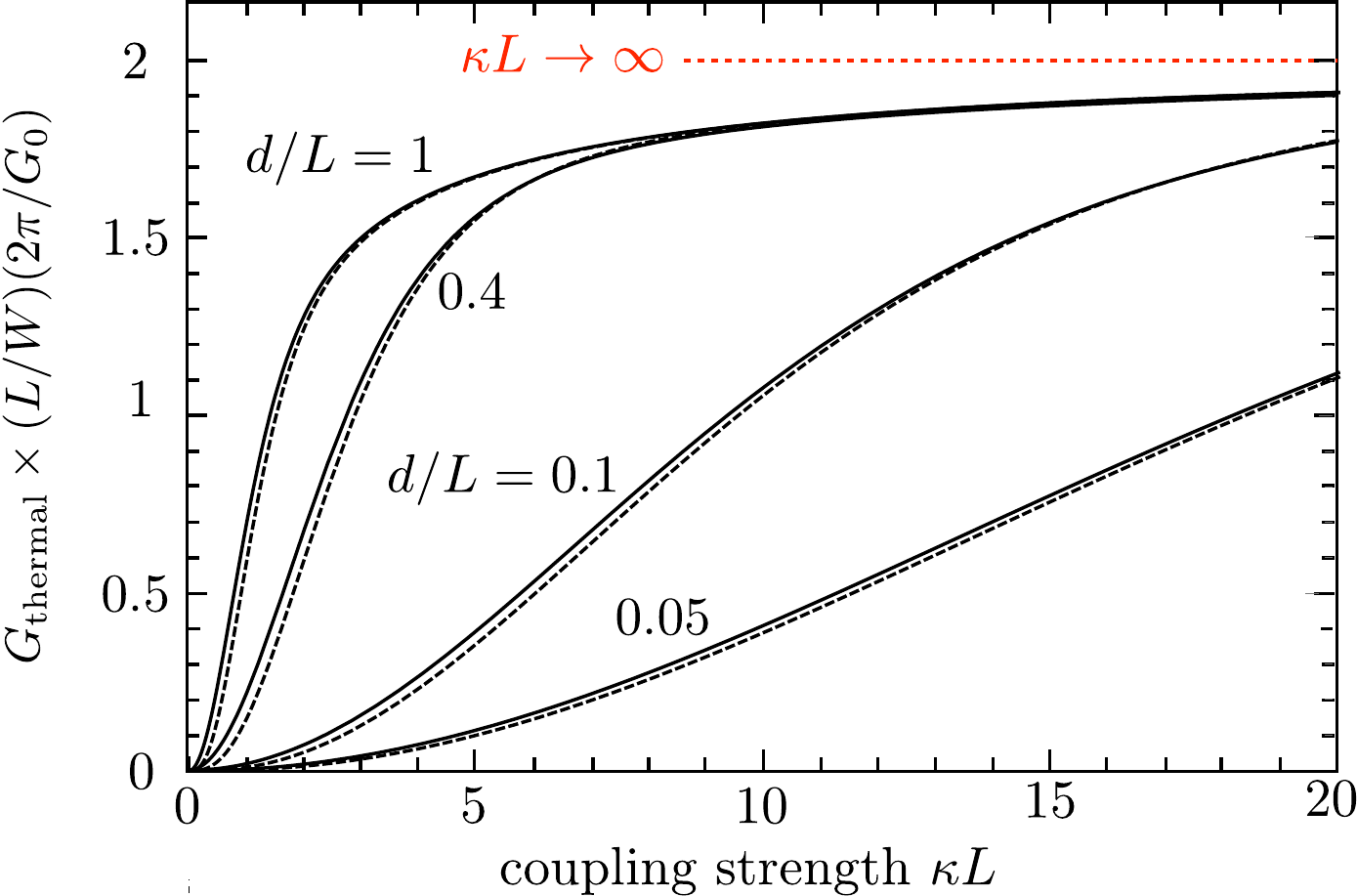}}
\caption{Thermal conductance for coupled top and bottom surfaces. For $d/L\gg 1$ we recover twice the single-surface plot in Fig.\ \ref{fig_Gthermal}. The solid curves for finite $d$ are calculated from Eq.\ \eqref{Gthermalresult2}, the dashed curves are the approximation \eqref{sigmathermalappr2}.
}
\label{fig_Gthermalcoupled}
\end{figure}

If we allow for coupling of the top and bottom surfaces via the metal contact, we would use the Green's function \eqref{Gkappagammacoupled} instead of Eq.\ \eqref{Gkappagamma}, to arrive at
\begin{align}
&G_{\rm thermal}=G_0\frac{W}{2\pi}\int_{0}^\infty dq\,2\kappa^2\sinh^2 \xi d\nonumber\\
&\quad\times(\xi\cosh qL\cosh \xi d+q\sinh qL\sinh \xi d)^{-2}.\label{Gthermalresult2}
\end{align}
There are now contributions from two surfaces in parallel, so $(L/W)G_{\rm thermal}\rightarrow 2\times G_{0}/2\pi$ in the large-$\kappa$ limit. As shown in Fig.\ \ref{fig_Gthermalcoupled}, the finite-$d$ effect is accurately described by a reduction factor $\tanh^2 \kappa d$,
\begin{equation}
G_{\rm thermal}\approx \frac{G_0 W}{2\pi L}\frac{2\kappa L\tanh^2\kappa d}{1+\kappa L}.\label{sigmathermalappr2}
\end{equation}

\subsection{Finite aspect ratio}
\label{sec_aspectratio}

Deviations from the universal limit \eqref{sigmathermaluniversal} of the thermal conductivity appear even for strong coupling to the metal contacts, if the aspect ratio of the surface area is not large enough. The relevant variable is the ratio $r={\cal P}/L$ of the perimeter ${\cal P}=2(W+d)$ of the metal contacts relative to their separation $L$. 

\begin{figure}[tb]
\centerline{\includegraphics[width=0.7\linewidth]{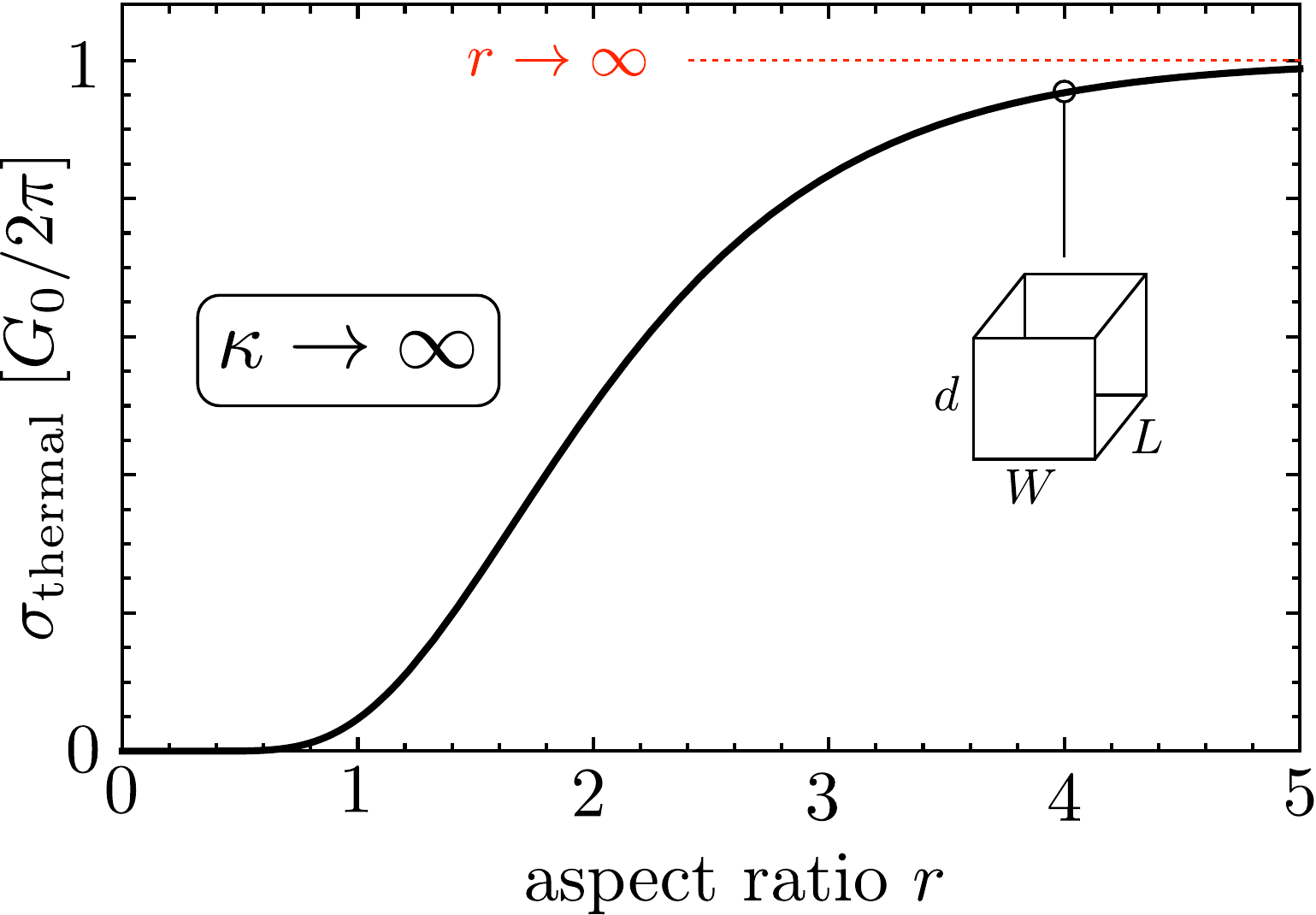}}
\caption{Thermal conductivity in the limit $\kappa\rightarrow\infty$ of strong coupling to the metal contacts, as a function of the aspect ratio $r=2(W+d)/L$. The curve is calculated from Eq.\ \eqref{sigmathermalfiniter}. A cube has $r=4$ and $\sigma_{\rm thermal}=0.95\times G_0/2\pi$.
}
\label{fig_aspectratio}
\end{figure}

Transverse momenta are quantized with antiperiodic boundary conditions, because of the $\pi$ Berry phase,
\begin{equation}
q_n=(2n+1)\pi/{\cal P},\;\;n=0,\pm 1,\pm 2,\ldots.\label{qndef}
\end{equation}
For strong coupling to the contacts ($\kappa \gg 1/L,1/d$) we find from Eq.\ \eqref{Gthermalresult} the thermal conductivity $\sigma_{\rm thermal}=(L/{\cal P})G_{\rm thermal}$ as the sum
\begin{equation}
\sigma_{\rm thermal}=G_0\sum_{n=-\infty}^\infty\frac{1}{2r\cosh^{2}[(2n+1)\pi/r]}.\label{sigmathermalfiniter}
\end{equation}
As shown in Fig.\ \ref{fig_aspectratio}, the universal limit is reached rather quickly; for a cube geometry, $L=W=d\Rightarrow r=4$, we are only 5\% below the universal limit.

\subsection{Locality condition}
\label{sec_locality}

In this model calculation the general locality condition \eqref{blockdiagonal} on the coupling matrix reads
\begin{equation}
w(z,q)\tau_z w^\dagger(z,q')=\delta_{qq'}R(z),\label{WzqWzqdaggermodel}
\end{equation}
since $\delta_{qq'}\mapsto \delta(x-x')$ upon Fourier transformation. The $2\times 2$ matrix $R$, acting on the spin degree of freedom, may depend on $z$ but it should not depend on $q$. We only need to impose locality in $x$, because in Eq.\ \eqref{SWGdef} we have already taken a local coupling in $z$.

The electron-hole and time-reversal symmetry constraints
\begin{equation}
w(z,q)=w^\ast(z,-q)\tau_x,\;\;w(z,q)=\sigma_y w^\ast(z,-q)\sigma_y\label{wzqsymmetry}
\end{equation}
 require that
\begin{equation}
R(z)=-R^{\rm T}(z)=-\sigma_y R(z)\sigma_y,\label{Rsymmrelation}
\end{equation}
and this is only possible for a $2\times 2$ matrix\cite{note3} if $R(z)\equiv 0$. Then Eq.\ \eqref{tWG} gives $t^\dagger(z,z')\tau_z t(z,z')\equiv 0$ and thus Eq.\ \eqref{PG0ehsymm} implies the equality 
\begin{equation}
P_{\rm shot}/P_0=G_{\rm thermal}/G_0.\label{PGequality2D}
\end{equation}
All the results presented above for the thermal conductance then apply also to the shot noise power.

Within the 2D model calculation of this section we cannot ascertain that the locality condition \eqref{WzqWzqdaggermodel} holds. For that purpose we need to perform a fully 3D calculation, as we will do in the next section.

\section{Numerical solution of the full 3D scattering problem}
\label{sec_3D}

\subsection{Model Hamiltonian}
\label{3D_modelH}

Our numerical simulation is based on the Bogoliubov-De Gennes Hamiltonian\cite{Qi11} 
\begin{align} 
&H({\bf k})= \left( \frac{{\bf k}^2}{2m} +V(\bm{r})- E_{\rm F} \right) \sigma_0 \otimes \tau_z\nonumber\\
&\quad + \Delta \left( k_z  \sigma_z \otimes \tau_x  - k_y  \sigma_0 \otimes \tau_y - k_x  \sigma_x  \otimes \tau_x \right), \label{H_k}
\end{align}
discretized on a cubic lattice (lattice constant $a_0$, hopping energy $t_0$). The disorder potential is $V$ and the \textit{p}-wave pair potential is $\Delta$. This is a generic model of a 3D topological superconductor in symmetry class DIII, without spin-rotation symmetry but with electron-hole and time-reversal symmetries
\begin{equation}
H(\bm{k})=-\tau_x H^\ast(-\bm{k})\tau_x,\;\;H(\bm{k})=\sigma_y H^\ast(-\bm{k})\sigma_y.\label{PH}
\end{equation}

The geometry is that of Fig.\ \ref{fig_layout}, in the normal-metal regions we set $\Delta\equiv 0$. A tunnel barrier of height $U_{\rm barrier}$ (two lattice sites wide) is introduced at the NS interfaces $z=\pm L/2$. The scattering matrix is calculated using the Kwant toolbox.\cite{kwant}  We fixed the Fermi energy at $E_{\rm F}=2.5\,t_0$ and took a relatively large pair potential $\Delta=0.4\,E_{\rm F}$ to eliminate bulk conduction without requiring a large $L$. 

\subsection{Translationally invariant system}
\label{3D_translinv}

\begin{figure}[tb]
\centerline{\includegraphics[width=0.8\linewidth]{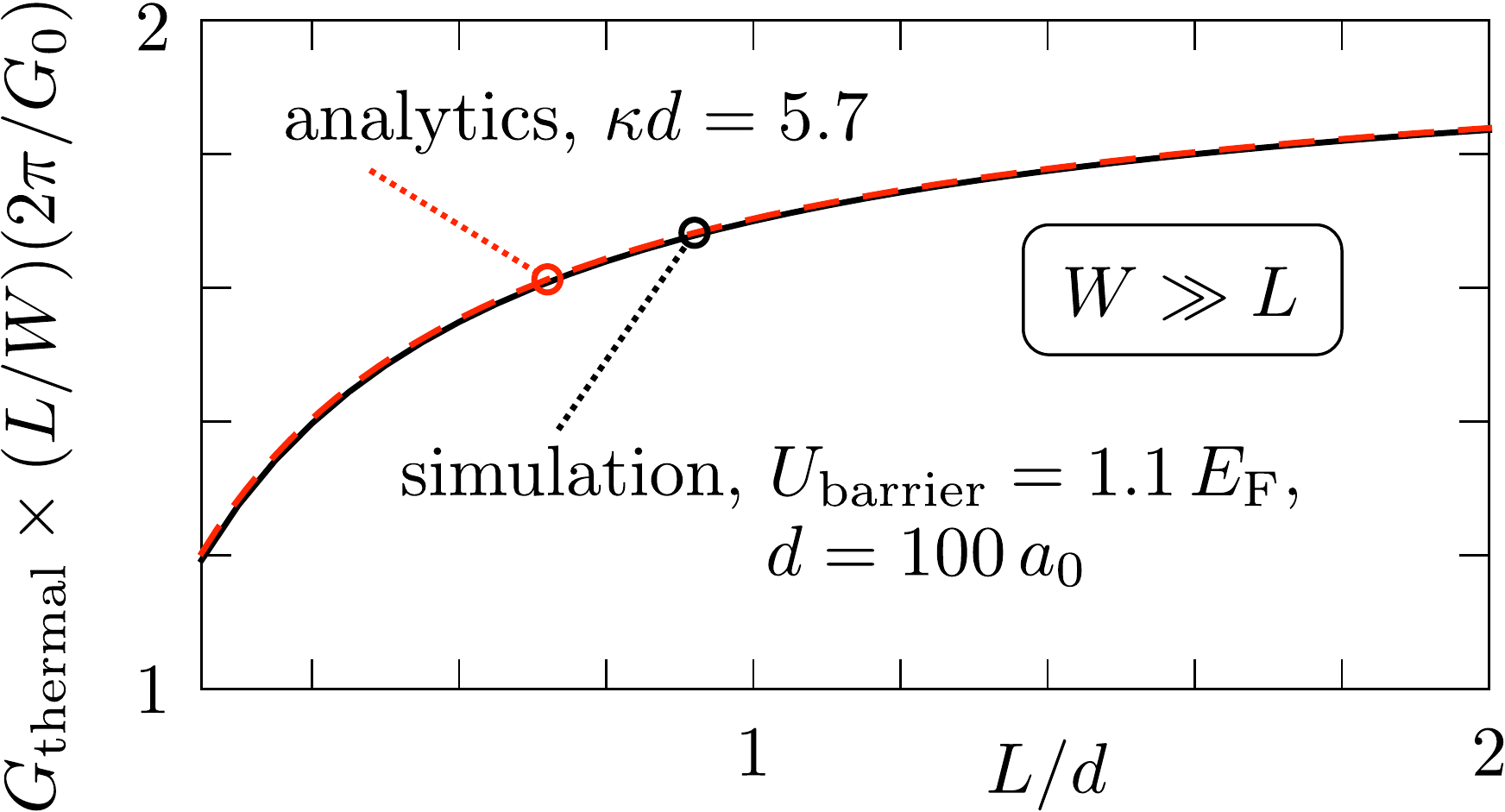}}
\caption{Comparison of the analytical result \eqref{Gthermalresult2} for the thermal conductance with a numerical simulation of the Hamiltonian \eqref{H_k}. The coupling strength $\kappa$ at the NS interface is the single fit parameter for the comparison. These are calculations in the simplest case $V_0=0$,  $B_0=0$, $\alpha_{\rm so}=0$, $W/L\rightarrow\infty$.
}
\label{fig_widenumerics1}
\end{figure}

For a direct test of the analytical calculation from the previous section we first consider a translationally invariant system along the $x$-direction ($W\gg L,d\simeq 100a_0$, no disorder). Results are shown in Fig.\ \ref{fig_widenumerics1}. The analytical result Eq.\ \eqref{Gthermalresult2} describes the numerics very well, with the coupling strength $\kappa$ as the single fit parameter. This demonstrates the validity of the 2D representation of the 3D scattering problem, including the effect of coupling between top and bottom surfaces.

\begin{figure}[tb]
\centerline{\includegraphics[width=0.8\linewidth]{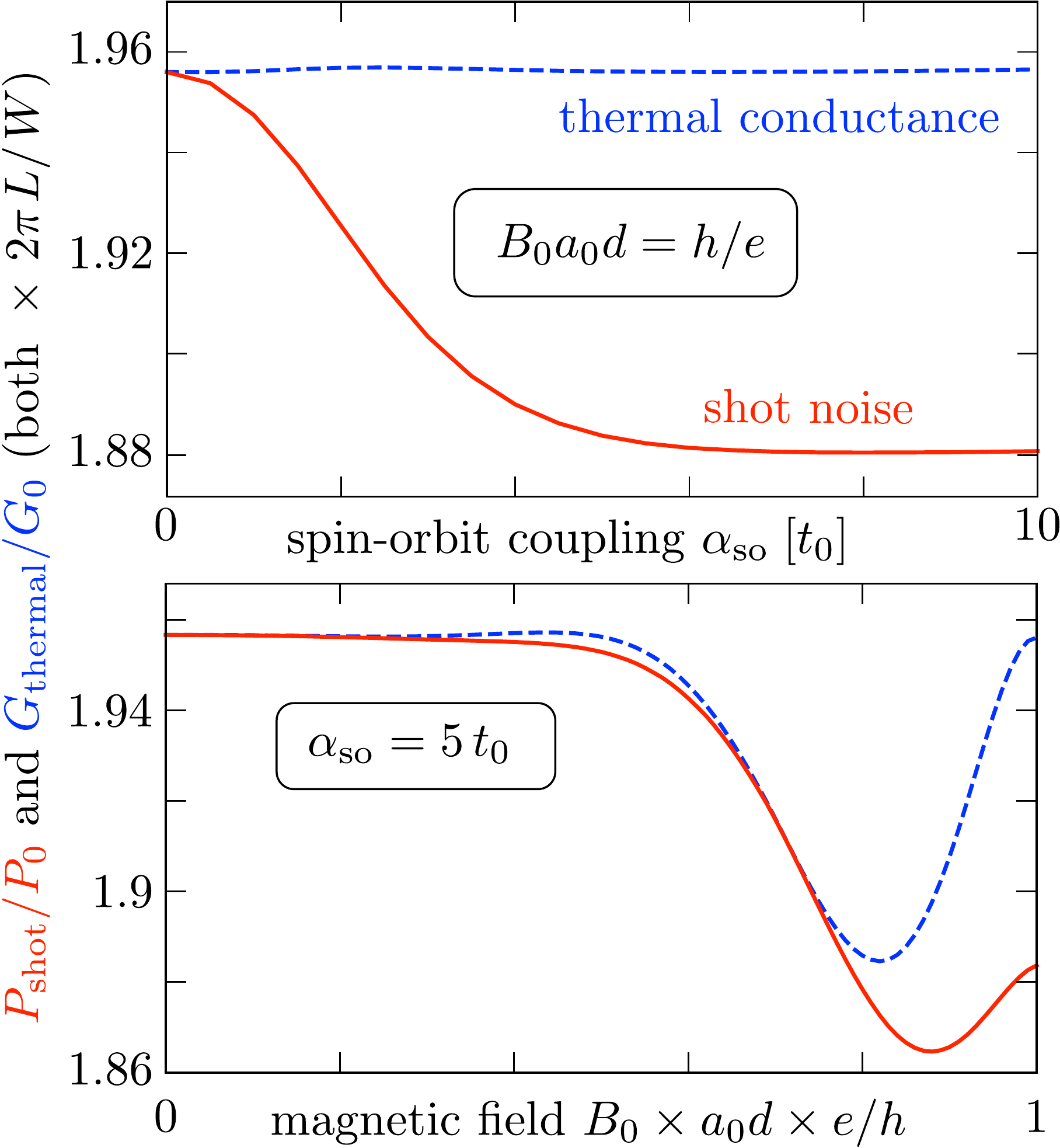}}
\caption{Numerical results for the shot noise power (red solid curves) and  thermal conductance (blue dashed curves), in a superconductor with $L=d=100\,a_0$, $W/L\rightarrow\infty$, $V_0=0$, $U_{\rm barrier}=0$. The magnetic field and spin-orbit coupling are varied in, respectively, the bottom and top panel. This simulation demonstrates the inequality $P_{\rm shot}/P_0\leq G_{\rm thermal}/G_0$, with equality if either $B_0$ or $\alpha_{\rm so}$ vanishes.
}
\label{fig_widenumerics2}
\end{figure}

We next investigate the extent to which the relation \eqref{PGequality2D} holds, still in the translationally invariant system, by adding to the Hamiltonian \eqref{H_k} the spin-orbit coupling
\begin{equation}
V_{\rm so}=\alpha_{\rm so}k_x\sigma_y\otimes\tau_z,\label{Vsodef}
\end{equation}
in order to mix the modes from top and bottom surface. (The same $V_{\rm so}$ is added to superconducting and normal regions.) Note that $V_{\rm so}$ preserves the electron-hole and time-reversal symmetries \eqref{PH}. We break time-reversal symmetry by imposing on the normal metal the magnetic field $\bm{B}=B_0\theta(|z|-L/2)\hat{x}$, in the gauge $\bm{A}=B_0 \theta(|z|-L/2)\,y\hat{z}$. 

As shown in Fig.\ \ref{fig_widenumerics2}, both a nonzero $\alpha_{\rm so}$ and a nonzero $B_0$ are needed for a difference between dimensionless shot noise power and thermal conductance. The nonzero $\alpha_{\rm so}$ is needed to couple the modes from top and bottom surface --- otherwise the transmission matrix would be of rank one and the equality \eqref{PGequality2D} would hold irrespective of whether time-reversal symmetry is broken or not.\cite{Akh11} The nonzero $B_0$ is needed because of the argument from Sec.\ \ref{conditiononXi} that mode coupling in the presence of time-reversal symmetry is not effective at violating the relation between shot noise power and thermal conductance.
 
\subsection{Disorder effects}
\label{3D_disorder}

\begin{figure}[tb]
\centerline{\includegraphics[width=0.8\linewidth]{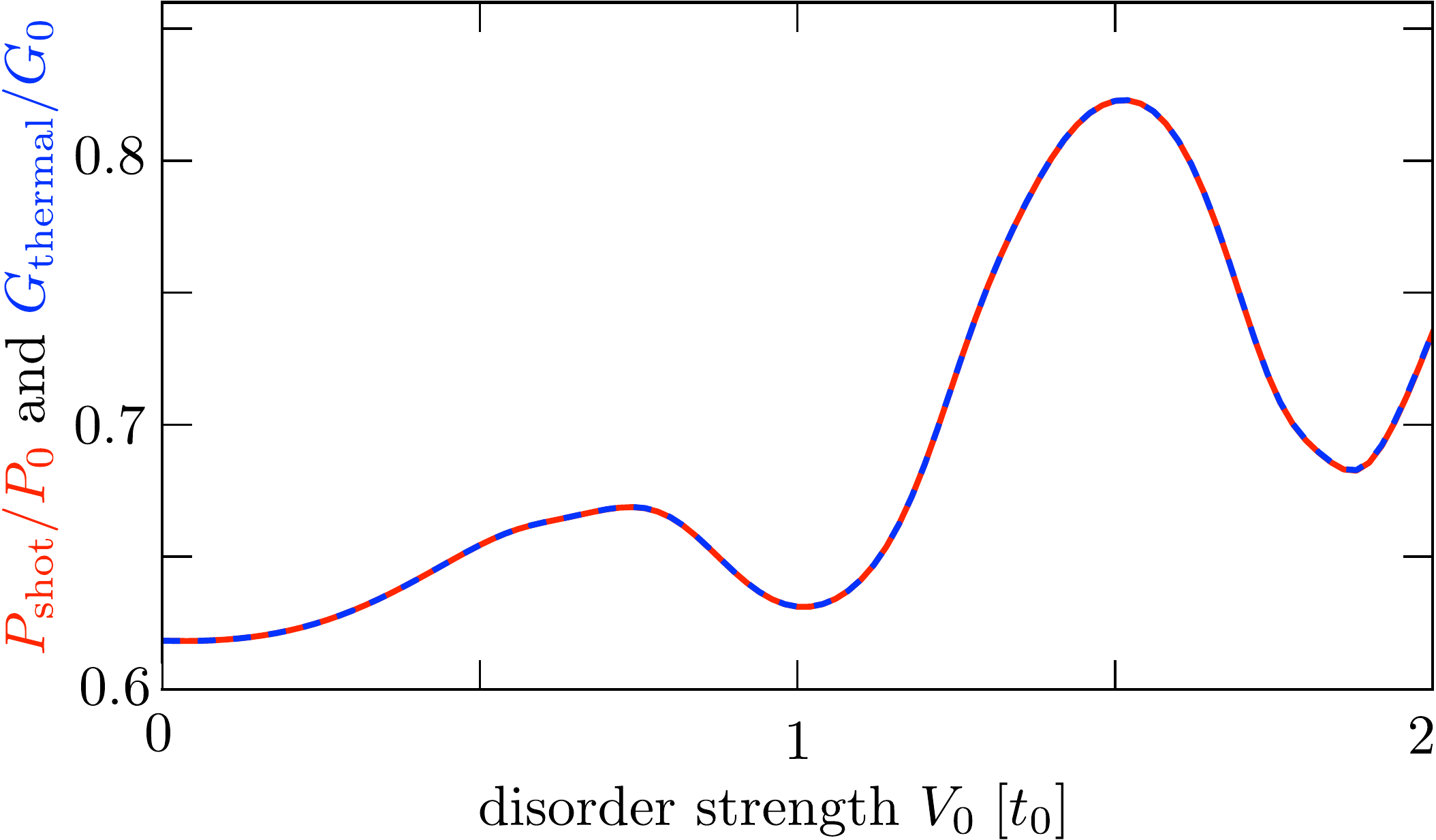}}
\caption{Numerical results for the shot noise power (red solid curve) and thermal conductance (blue dashed curve) as a function of the disorder potential strength $V_0$, in a superconductor with $L=d=W=20\,a_0$ ($B_0=0$, $\alpha_{\rm so}=0$, $U_{\rm barrier}=0$). Shot noise and thermal conductance differ by less than $10^{-3}$, even in the presence of significant disorder.
}
\label{fig_cubenumerics}
\end{figure}

We now break translational invariance by adding a disorder potential $V$, uniformly distributed in the interval $(-V_0,+V_0)$, randomly fluctuating from site to site throughout the superconductor. We also added disorder on the normal side of the NS interface (in a sheet of width $10a_0$). Because the calculations are now computationally more expensive we took a smaller superconductor, a cube of size $L=W=d=20a_0$. Results are shown in Fig.\ \ref{fig_cubenumerics}.

Without disorder the thermal conductivity is close to the limit expected from Fig.\ \ref{fig_Gthermalcoupled} for a cube aspect ratio: $\sigma_{\rm thermal}/G_0=0.95/2\pi\Rightarrow G_{\rm thermal}/G_0=0.605$. Disorder has a significant effect, but the dimensionless shot noise and thermal conductance remain nearly indistinguishable.

\section{Discussion}
\label{conclude}

As a particle that is its own antiparticle, a Majorana fermion must be charge-neutral --- but it need not be in an eigenstate of charge. This is a key distinction between a Majorana fermion as a fundamental particle such as a neutrino, or as a composite quasiparticle in a superconductor. For the latter only the expectation value of the charge must vanish, so there may be quantum fluctuations of the charge. Here we have shown how we can exploit this property for a purely electrical detection of the Majorana surface states in a 3D topological superconductor, obviating the difficulty of thermal measurements.

We like to think of the relation
\begin{equation}
G_{\rm thermal}/P_{\rm shot}={\cal L}T/eV,\;\;{\cal L}\equiv\tfrac{1}{3}(\pi k_{\rm B}/e)^2,\label{WiedemannFranzM}
\end{equation}
between thermal conductance and electrical shot noise power as the Majorana counterpart of the electronic Wiedemann-Franz relation
\begin{equation}
G_{\rm thermal}/G_{\rm electrical}={\cal L}T\label{WiedemannFranz}
\end{equation}
between thermal and electrical conductance.\cite{Nom12} The analogy is quite direct: Eq.\ \eqref{WiedemannFranz} expresses the fact that a nonequilibrium electron transports energy and charge in a fixed ratio. The same holds for Eq.\ \eqref{WiedemannFranzM}, with the electron charge $Q=e$ replaced by the Majorana charge variance ${\rm Var}\,Q=e^2$.

There exists an altogether different ``Wiedemann-Franz type relation'' for Majorana fermions, relating heat and \textit{particle} currents rather than heat and \textit{charge} currents:
\begin{equation}
G_{\rm thermal}/G_{\rm particle}=\tfrac{1}{2}G_0,\;\;G_0=\tfrac{1}{3}(\pi k_{\rm B})^2 \,T.\label{WiedemannFranz2}
\end{equation}
The ``particle conductance'' $G_{\rm particle}$ is not directly measurable (since Majorana fermions do not couple to the chemical potential), but it can be formally defined in terms of the Landauer formula $G_{\rm particle}=N{\cal T}/h$ or in terms of an equivalent Kubo formula.\cite{Nom12} The factor $1/2$ is the ``topological'' or ``central'' charge ${\cal C}=1/2$ of a Majorana fermion.\cite{Rea00} No such factor appears in Eq.\ \eqref{WiedemannFranzM}, because both the thermal conductance and the shot noise power are proportional to ${\cal C}$, so it drops out of the ratio.

One direction for future research is to generalize the relation \eqref{WiedemannFranzM} to topological superconductors with more than a single species of Majorana fermions on their surface. This is a key difference with 3D topological insulators, which have a $\mathbb{Z}_2$ topological quantum number, so at most a single Dirac cone on the surface. In contrast, 3D topological superconductors have a $\mathbb{Z}$ topological quantum number, allowing for multiple Majorana cones.\cite{Has10,Qi11,Ryu10} 

Another direction to explore is how the class-DIII topological superconductors with Majorana surface states considered here compare with the class-CI topological superconductors with Dirac surface states. For Eq.\ \eqref{WiedemannFranz2} the difference is simply a factor of two, to account for a central charge ${\cal C}=1$ of Dirac fermions.\cite{Xie15} We do not expect such a simple correspondence for the relation \eqref{WiedemannFranzM}.

From the experimental point of view, the usefulness of Eq.\ \eqref{WiedemannFranzM} is that it provides a purely electrical way to access the transport properties of Majorana surface states. The shot noise measurements should be performed at energies $eV$ well below the superconducting gap $\Delta$. In Cu$_x$Bi$_2$Se$_3$ this is about 0.6~meV.\cite{Sas11} Shot noise dominates over thermal noise if $eV\gtrsim 3k_{\rm B}T$,\cite{Gne15} so if one would perform the experiment at $V=0.1\,{\rm meV}$, a readily accessible temperature range $T\lesssim 0.3\,{\rm K}$ would do.

\acknowledgments
This research was supported by the Foundation for Fundamental Research on Matter (FOM), the Netherlands Organization for Scientific Research (NWO/OCW), and an ERC Synergy Grant.

\appendix
\section{Calculation of the matrix Green's function of the surface Hamiltonian}
\label{app_G}

We solve the differential equation
\begin{align}
\left[-i\sigma_x\frac{\partial}{\partial x} +  q\sigma_z -\tfrac{1}{2}i\Gamma(z)\sigma_0-E\right]{\cal G}(z,z';q,E)\nonumber\\
=\openone \delta(z-z')\label{Greenequation}
\end{align}
to obtain the $2\times 2$ matrix Green's function that determines the transmission matrix of the Majorana fermions between the normal-metal contacts. (To simplify the notation we have set $v\equiv 1$.) For a similar calculation in graphene, see Ref.\ \onlinecite{Tit10}.

We first consider in Sec.\ \ref{app_Gsingle} the case of uncoupled top and bottom surfaces, when the $z$-coordinate ranges over the entire real axis and the tunnel coupling in the contact region is given by
\begin{equation}
\Gamma(z)=\Gamma\theta(|z|-L/2),\;\;-\infty<z<\infty.\label{Gammazsingle}
\end{equation}
In Sec.\ \ref{app_Gdouble} we incorporate the finite extension $d$ of the contact region, by setting
\begin{equation}
\Gamma(z)=\Gamma\left[\theta(|z|-L/2)-\theta(|z|-L/2-d)\right],\;\;|z|<L+d.\label{Gammazcoupled2}
\end{equation} 
Antiperiodic boundary conditions at $z=\pm(L+d)$ then couple the top and bottom surfaces.

\begin{widetext}

\subsection{Single surface}
\label{app_Gsingle}

We define
\begin{equation}
\varepsilon(z)=E+\tfrac{1}{2}i\Gamma(z),\;\;\varepsilon_0=E+\tfrac{1}{2}i\Gamma,\label{varepsilondef}
\end{equation}
with $\Gamma(z)$ given by Eq.\ \eqref{Gammazsingle}. The Green's function that decays at infinity is

\begin{align}
{\cal G}(z,z';q,E)={}&{\cal P}_{z\leftrightarrow 0}\exp\left[\int_{0}^{z}dz_1\left( i\sigma_z\varepsilon(z_1)+q\sigma_y\right)\right][M-\tfrac{1}{2}i\sigma_z+i\sigma_z\theta(z-z')] \nonumber\\
&\cdot{\cal P}_{0\leftrightarrow z'}\exp\left[\int_{0}^{z'}dz_2 \left( -i\sigma_z\varepsilon(z_2)+q\sigma_y\right)\right],\label{Ginfinity}
\end{align}
where ${\cal P}_{z_1\leftrightarrow z_2}$ indicates a monotically increasing or decreasing ordering of the $z$-dependent non-commuting operators, from $z_1$ leftmost to $z_2$ rightmost.  The matrix $M$ is determined by the requirement that $\lim_{z\rightarrow\pm\infty} {\cal G}(z,z';q,E)=0$:
\begin{align}
&\left(-\varepsilon_0\pm i\sqrt{q^2-\varepsilon_0^2},q\right)\cdot\exp[\pm \tfrac{1}{2}L(iE\sigma_z+q\sigma_y)](M\pm \tfrac{1}{2}i\sigma_z)=0.\label{Mrequirement}
\end{align}
The row-spinor on the left-hand-side is orthogonal (without taking complex conjugates) to the column-spinor
\begin{equation}
|\pm\rangle=\begin{pmatrix}
\varepsilon_0\mp i\sqrt{q^2-\varepsilon_0^2}\\
q\
\end{pmatrix},
\end{equation}
which is an eigenstate of
\begin{equation}
(i\varepsilon_0\sigma_z+q\sigma_y)|\pm\rangle=\pm\sqrt{q^2-\varepsilon_0^2}|\pm\rangle.
\end{equation}
The square root should be taken such that ${\rm Re}\,\sqrt{q^2-\varepsilon^2}>0$. 

The result is
\begin{equation}
\begin{split}
&M=\begin{pmatrix}
M_{1}&M_{2}\\
M_{2}&M_{1}
\end{pmatrix},\;\;
M_1=\frac{E(\xi_0^2-\tfrac{1}{2}i\Gamma E)\cosh(L\xi_0)+E\xi\xi_0\sinh(L\xi_0)+\tfrac{1}{2}i\Gamma q^2}{2\xi\xi_0^2\cosh(L\xi_0)+2(\xi_0^2-\tfrac{1}{2}i\Gamma E)\xi_0\sinh(L\xi_0)},\\
&M_2=\frac{q(\xi_0^2-\tfrac{1}{2}i\Gamma E)\cosh(L\xi_0)+q\xi\xi_0\sinh(L\xi_0)+\tfrac{1}{2}i\Gamma qE}{2\xi\xi_0^2\cosh(L\xi_0)+2(\xi_0^2-\tfrac{1}{2}i\Gamma E)\xi_0\sinh(L\xi_0)},
\end{split}
\end{equation}
with the definitions
\begin{equation}
\xi_0=\sqrt{q^2-E^2},\;\;\xi=\sqrt{q^2-(E+\tfrac{1}{2}i\Gamma)^2}.
\end{equation}
As a check, we take the limit $E\rightarrow 0$, $q\rightarrow 0$, when $M\rightarrow \tfrac{1}{2}i\sigma_0$, as it should. Note the symmetry relations
\begin{equation}
\sigma_y M^{\rm T}(-q,E)\sigma_y=M(q,E),\;\;M^\ast(-q,-E)=-M(q,E),
\end{equation}
which ensure that the Green's function \eqref{Ginfinity} satisfies the required time-reversal and electron-hole symmetries:
\begin{equation}
\sigma_y{\cal G}^{\rm T}(z',z;-q,E)\sigma_y={\cal G}(z,z';q,E),\;\;{\cal G}^\ast(z,z';-q,-E)=-{\cal G}(z,z';q,E).\label{Gsymmetries}
\end{equation}

To obtain the transmission matrix we set $z>L/2$ and $z'<-L/2$:
\begin{align}
{\cal G}(z,z';q,E)={}&\exp\left[(z-L/2)\left( i E\sigma_z-\tfrac{1}{2}\Gamma\sigma_z+q\sigma_y\right)\right]
\exp\left[(L/2)\left( iE\sigma_z+q\sigma_y\right)\right]\nonumber\\
&\cdot \left(M+ \tfrac{1}{2}i\sigma_z\right)\exp\left[(L/2)\left( iE\sigma_z-q\sigma_y\right)\right]\exp\left[-(z'+L/2)\left( iE\sigma_z-\tfrac{1}{2}\Gamma\sigma_z-q\sigma_y\right)\right].\label{GsingleresultApp}
\end{align}
For $E=0$ this simplifies to Eq.\ \eqref{Gkappagamma} in the main text.

\subsection{Coupled top and bottom surfaces}
\label{app_Gdouble}

The Green's function for coupled top and bottom surfaces is still of the form \eqref{Ginfinity}, with $\Gamma(z)$ now given by Eq.\ \eqref{Gammazcoupled2}. Instead of a decay at infinity we now have the antiperiodic boundary conditions
\begin{equation}
{\cal G}\left(L+d, z';q,E\right)= - {\cal G}\left(-L-d, z';q,E\right). \label{pbc}
\end{equation}
The condition \eqref{Mrequirement} on the matrix $M$ is replaced by
\begin{align}
&\exp\left[ \tfrac{1}{2}L\left(iE\sigma_z+q \sigma_y\right)\right]  \exp\left[d\left(i\varepsilon_0\sigma_z+ q \sigma_y\right)\right] \exp\left[\tfrac{1}{2}L\left(iE\sigma_z+ q \sigma_y\right)\right]  \left( M +\tfrac{1}{2}i \sigma_z\right) \nonumber \\ 
 &\quad = -  \exp\left[ - \tfrac{1}{2}L\left(iE\sigma_z+q \sigma_y\right)\right]  \exp\left[ - d\left(i\varepsilon_0\sigma_z+ q \sigma_y\right)\right] \exp\left[ - \tfrac{1}{2}L\left(iE\sigma_z+ q \sigma_y\right)\right]  \left( M - \tfrac{1}{2}i \sigma_z\right), \label{pbcE_M}
\end{align}
with solution
\begin{equation}
\begin{split}
&M=\begin{pmatrix}
M_{1}&M_{2}\\
M_{2}&M_{1}
\end{pmatrix},\;\;
M_1=\frac{E(\xi_0^2-\tfrac{1}{2}i\Gamma E)\cosh(L\xi_0)\tanh(\xi d)+E\xi\xi_0\sinh(L\xi_0)+\tfrac{1}{2}i\Gamma q^2\tanh(\xi d)}{2\xi\xi_0^2\cosh(L\xi_0)+2(\xi_0^2-\tfrac{1}{2}i\Gamma E)\xi_0\sinh(L\xi_0)\tanh(\xi d)},\\
&M_2=\frac{q(\xi_0^2-\tfrac{1}{2}i\Gamma E)\cosh(L\xi_0)\tanh(\xi d)+q\xi\xi_0\sinh(L\xi_0)+\tfrac{1}{2}i\Gamma qE\tanh(\xi d)}{2\xi\xi_0^2\cosh(L\xi_0)+2(\xi_0^2-\tfrac{1}{2}i\Gamma E)\xi_0\sinh(L\xi_0)\tanh(\xi d)}.
\end{split}\label{McoupledApp}
\end{equation}

For the transmission matrix we set $-L/2-d<z'<-L/2$, $L/2<z<L/2+d$. The energy-dependent Green's function is then given by Eq.\ \eqref{GsingleresultApp} with $M$ from Eq.\ \eqref{McoupledApp}. At zero energy this produces the result \eqref{Gkappagammacoupled} from the main text.

\end{widetext}


\begin{thebibliography}{99}
\bibitem{note0} Symmetry class DIII refers to a superconductor with broken spin-rotation symmetry. Class-CI superconductors with preserved spin-rotation symmetry may also host gapless surface states, but these are Dirac fermions rather than Majorana fermions and do not have the same protection against disorder scattering.
\bibitem{note2} The symmetry operations \eqref{SehTsym} on the scattering matrix are taken in a time-reversally symmetric basis of the propagating modes in the normal metal. If transverse momentum $q$ is used to label the modes, these operations should be accompanied by $q\mapsto -q$.
\bibitem{note1} The factor of two difference between the number of independent modes in the normal metal and the superconductor is the reason that the mode-matching approach of Ref.\ \onlinecite{Two06} cannot be applied here.
\bibitem{note3} Any $2\times 2$ antisymmetric matrix $R$ is a scalar multiple of $\sigma_y$, so if we insist that $R$ anticommutes with $\sigma_y$ we must have $R\equiv 0$.
\end{thebibliography}
\end{document}